\documentstyle[12pt,damtp]{damtpprint}
\pubdate{July 1991}
\pubnumber{DAMTP-91/27}
\title{Signature Characters for the Virasoro Algebra}
\author{Adrian Kent\damtpitp}
\abstract{
The number of ghost states at each energy level in a non-unitary conformal
field theory is encoded in the signature characters of the relevant
Virasoro algebra highest weight representations.
We give expressions for these signature characters.
These results complete Friedan-Qiu-Shenker's analysis of
the Virasoro algebra's highest weight representations.
}

\begin{document}
\maketitle

\input mssymb
\newenvironment{proof}{{\bf Proof}}{}
\newtheorem{lemma}{Lemma}
\newtheorem{theorem}[lemma]{Theorem}
\newtheorem{corollary}[lemma]{Corollary}
\def\floor#1{\lfloor#1\rfloor}
\def\blank#1{}
\def\swap#1#2{#1}
\def \ul {\underline}
\def \eps {\epsilon}
\def\ut#1{\hbox{\boldmath #1}}
\def\nn{\nonumber\\}
\def\tr{{\rm tr}}
\def\fif{{\rm~if~}}
\def\num{{\rm num}}
\def\sgn{{\rm sgn}}
\def\sig{{\rm sig}}
\def\diag{{\rm diag}}
\def\for{{\rm ~for~}}
\def\and{{\rm ~and~}}
\def\pol{{\rm pol}}
\def\quart{\frac{1}{4}}
\def\half#1{\frac{#1}{2}}
\def \hb {h_\beta}
\def \ha {h_\alpha}
\def \hbd {$\hb$}
\def \had {$\ha$}

By definition, the canonical bilinear form on the state space of a non-unitary
conformal field theory has indefinite signature.
This splits the space into subspaces on which the form is positive
and negative definite.
Since the state spaces are built from
Virasoro algebra highest weight representations, this
decomposition can be described in terms of the
generating functions which encode the signatures of the
inner product matrices on these representations.
Here we give expressions for these
generating functions for all the physically relevant (that is, highest
weight) representations of the Virasoro algebra.

The only useful results in this direction were obtained by Friedan,
Qiu and Shenker (FQS) \cite{fqs}, who analysed the dependence of a
representation's inner product on the representation's location
amongst the vanishing curves defined by the Kac determinant formula
\cite{kac,ff}.  For almost all representations for which the central
charge $c$ is less than one, FQS gave a level at which the inner
product matrix is not positive definite.  Using evidence from statistical
physics and elsewhere, FQS conjectured that the remaining representations,
which fall into a discrete series, are unitary.  Goddard, Kent and
Olive (GKO)\cite{gko} proved that these representations are indeed unitary.
Our results generalise those of FQS and GKO.

Recall that the Virasoro algebra has commutation relations
\eq
\begin{array}{rcl}
{}~[L_m , L_n ] \, \, & = & (m - n ) L_{m+n} + \frac{m^3 - m}{12}
\delta_{m,-n} C \, , \\ ~[L_m , C ] \, \, & = & 0 \, .
\end{array}
\en
The highest weight representation $W(h,c)$ (where $h$ and $c$ are
numbers, which we take to be real) is the irreducible representation which
contains a vector $\ket{h}$ such that
\eq
\begin{array}{rcll}
L_m \ket{ h } &=& 0 & {\rm~if~} m>0 \, , \\ L_0 \ket{ h } &=& h
\ket{h} \, , & \\ C \ket{h} &=& c \ket{h} \, , &
\end{array}
\en
and which is spanned by states $L_{-i_1} \ldots L_{-i_r} \ket{h}$ with
the $i_j > 0$.
We have the decomposition
\eq
W(h,c) = \bigoplus_{n = 0,1,2 \ldots} ( W(h,c) )_{n} \, ,
\en
where
$W(h,c)_{n}$ is the eigenspace of $L_0$ with eigenvalue $(h+n)$.
The canonical bilinear form  on $W(h,c)$ is defined by taking
$\ipone{h}{h} = 1$, setting $(L_n)^{\dagger} = L_{-n}$ and $C^{\dagger} = C$.
We write $M_{n}(h,c)$ for the real symmetric matrix
defining (in some choice of
basis) the inner product restricted to $W(h,c)_{n}$.

Now the {\it normalised signature character} of $W(h,c)$ is defined as
\eq
\sigma (h,c) =
\sum_{n = 0,1,2 \ldots} {\rm sig} (M_{n} (h,c ) ) t^n \, ,
\en
where ${\rm sig}(M)$ denotes the signature of the matrix $M$; that is,
if $M = S D S^T$, where $S$ is non-singular and $D = {\rm diag}(+1,
\ldots, +1, 0, \ldots , 0, -1 , \ldots , -1 )$, then ${\rm sig}(M) =
{\rm tr} (D)$.

We recall also the Kac determinant formula,\cite{kac,ff} which states that
\eq
\label{kac}
\det( M_n (h,c)) = \prod_{p q \leq n} ( h - h_{p,q} (c) )^{P(n-pq)} \, ,
\en
where if
\eq
c= c(m) \equiv 1 - \frac{6}{m(m+1)}
\en
then
\eq
h_{p,q} (c) = \frac{(( m+1)p - m q )^2 - 1 }{4 m (m+1)}
\en
and where
\eq
\prod_{n=1}^{\infty} ( 1 - t^n )^{-1} = \sum_{n= 0}^{\infty} P(n) t^n \, .
\en
Denote the vanishing curve $h= h_{p,q} (c)$ by $c_{p,q}$; call $pq$ the
{\it level} of $c_{p,q}$.
The Kac formula (\ref{kac}) implies that
if $R$ is a connected region of the $(h,c)$ plane,
containing no vanishing curves of level $\leq N$, then
$\sigma ( h,c)$ is constant to order $t^N$ on $R$;
in particular, if $R$ contains no vanishing curves, $\sigma$ is constant on
$R$.

Our first result is a difference equation for the signature characters.
We parametrize the regions $c<1$ and $c>25$ by $m>0$ and $-\frac{1}{2} < m < 0$
respectively.
For any irrational $m>-\frac{1}{2}$, we find that
\eq
\label{de1}
\begin{array}{l}
 \lim_{\delta \rightarrow 0^+} ((\sigma(h_{p,q} (m) + \delta, c(m) )
                              -(\sigma(h_{p,q} (m) - \delta, c(m) )) =   \\
{}~~~~~~~~~~~~~~~~~~~~~~~~~~~~~~~~~~
2 \eps(m,p,q) t^{pq} \sigma( h_{p,q} (m) + pq , c(m) ) \, ,
\end{array}
\en
where
\eq
\label{de2}
\eps(m,p,q) = \left\{
\begin{array}{ll}
(-1)^{\floor{\frac{p}{m}} + 1} & \fif m > 0 \and (m+1)p - m q < 0  \, , \\
(-1)^{\floor{\frac{q}{m+1}} } & \fif m > 0 \and (m+1)p - m q > 0  \, , \\
(-1)^{\floor{\frac{-p}{m}} + 1} & \fif -\frac{1}{2} < m < 0
                         \and (m+1)p + m q < 0  \, , \\
(-1)^{\floor{\frac{q}{m+1}} + 1} & \fif -\frac{1}{2} < m < 0
                         \and (m+1)p + m q > 0  \, .
\end{array}
\right.
\en

For $1<c<25$, the only vanishing curves in the real $(h,c)$ plane are the
lines $h = h_{p,p} (c)$, and we find that
\eq
\begin{array}{l}
{\displaystyle \lim_{\delta \rightarrow 0^+}}
((\sigma(h_{p,p} (c) + \delta, c )
                              -(\sigma(h_{p,p}(c) - \delta, c )) =  \\
{}~~~~~~~~~~~~~~~~~~~~~~~~~~~~~~~~~~~~~~~~
2 (-1)^{p+1} t^{p^2} \sigma( h_{p,p} (c) + p^2 , c )  \, .
\end{array}
\en

The matrices $M_n (h,c)$ become asymptotically diagonal as
$h \rightarrow \pm \infty$, and it is easy to show that
\eq
\label{bc}
\lim_{h \rightarrow  \infty} \sigma(h,c) =
\prod_{n=1}^{\infty} (1 - t^n )^{-1} \, ,
\en
for any $c$. Here, and elsewhere, the limit of a formal power series is
taken term by term: that is,
\eq
\lim_{h \rightarrow  \infty} \sigma(h,c) \equiv
\sum_{n=0,1,2, \ldots } \lim_{h \rightarrow \infty} \sig ( M_n (h,c )) t^n \, .
\en

If $(h,c(m))$ lies on no vanishing curves and
$m$ is irrational or non-real, then
$\sigma(h,c(m))$ is determined by
the difference equation (\ref{de1}) and the boundary
condition (\ref{bc}).
We find the remaining signature characters as follows.
If $m$ is rational and $(h,c(m))$ lies on no vanishing curve, then
we take a sequence $(m_i )_{i=1}^{\infty}$, converging to $m$,
such that the $m_i$ are irrational and the points $(h,c(m_i ))$ all lie on
no vanishing curve.
Then
\eq
\sigma ( h , c(m)) = \lim_{i \rightarrow \infty }  \sigma ( h , c(m_i )) \, .
\en
If $(h, c(m) )$ lies on precisely one vanishing curve, then
\eq
\sigma ( h , c) = \lim_{\delta \rightarrow 0^+ }
      \frac{1}{2} (\sigma(h + \delta, c ) + \sigma(h - \delta, c)) \, .
\en

Finally, if $(h,c(m))$ lies at an intersection of vanishing curves,
let $(p,q)$ be positive integers such that $h= h_{p,q}(c(m))$ and such that if
$h= h_{p',q'}(c(m))$ with $(p',q')$ positive integers then $pq \leq p'q'$.
Take sequences $(m_i^{+} )_{i=1}^{\infty}$, converging to $m$ from above,
and $(m_i^{-} )_{i=1}^{\infty}$, converging to $m$ from below,
such that the $m_i^{\pm}$ are irrational.
Then
\eq
\sigma ( h_{p,q}(m) , c(m)) = \frac{1}{2}(
\lim_{i \rightarrow \infty }  \sigma ( h_{p,q}(m_i^+ ) , c(m_i^+ )) +
\lim_{i \rightarrow \infty }  \sigma ( h_{p,q}(m_i^- ) , c(m_i^- ))) \, .
\en

These results imply simple limiting expressions for the signature characters.
For example, if $u \neq \frac{n^2 - 1}{24}$ for any positive integer $n$, then
we find
\eq
\label{asymp1}
\lim_{c \rightarrow \pm \infty} \sigma(- u c, c) =
\prod_{\{ n : u < \frac{n^2 - 1}{24} \}} ( 1 \mp t^n )^{-1}
\prod_{\{ n : u > \frac{n^2 - 1}{24} \}} ( 1 \pm t^n )^{-1}  \, .
\en
(These equations can also be obtained directly, since the matrices
$M_n (-u c, c )$ are diagonalised in the relevant limits.)
We also find that, if $m$ is a positive integer,
\eq
\lim_{c \rightarrow 1^-} \sigma ( \frac{m^2}{4}, c) =
(1 + 2 \sum_{r=1}^{\infty} (-1)^r t^{r(m+r)})
\prod_{n=1}^{\infty} (1 -t^n )^{-1}   \, .
\en

The signature characters in the region $c \geq 1$ are also relatively simple.
If $h>0$, then $W(h,c)$ is unitary and
$\sigma (h,c) = \prod_{n=1}^{\infty} ( 1 - t^n )^{-1}$.
If $h \leq 0$, the signature character takes the form
$p(t) \prod_{n=1}^{\infty} ( 1 - t^n )^{-1}$, where $p(t)$ is a polynomial.
For example, if $1 < c< 25$ and
$((p-1)^2 -1) (1-c ) > 24 h > (p^2 - 1) (1-c )$ for some positive
integer $p$, then
\eq
\sigma(h,c) = (1 + 2 \sum_{r=1}^{p-1} (-1)^r t^{r^2} )
\prod_{n=1}^{\infty}  (1 -t^n )^{-1}   \, .
\en

In the region $(c < 1, h > \frac{c-1}{24})$,
the solutions to the difference equation (\ref{de1}) are more complicated.
The most interesting (and physically relevant) representations are
those corresponding to rational $m$.
If $r$ and $s$ are coprime integers, $m= r/s>0$, we set
$h(a,m) = (a^2 - s^2 )/4 r(r+s)$.
We write
\eq
\eps^{\pm} (m,p,q) = \lim_{\delta \rightarrow 0^{\pm}}
                     \eps (m+\delta , p,q) \, .
\en
Then we find that, if $h(a-1,m) < h < h(a,m)$,
\eq
\label{gaps}
\begin{array}{rcl}
\sigma(h,c(m)) &=& {\displaystyle\prod_{n=1}^{\infty}} ( 1 - t^n )^{-1} \\
    && ( 1 + {\displaystyle\sum_{(p_1 , q_1 ), \ldots , (p_r , q_r )}} (-2)^r
                   \prod_{i=1}^r \eps^+ (m, p_i , q_i )
                    t^{ p_i q_i } ) \, ,
\end{array}
\en
where the sum is over all finite sequences of pairs of positive integers
such that
\begin{enumerate}
\item $| (r+s) p_1 - r q_1 | \geq a$,
\item for $i \geq 1$, $| (r+s) p_{i+1} - r q_{i+1} | > (r+s) p_{i} + r q_{i}$
or \newline $(r+s) p_{i+1} - r q_{i+1} = - ( (r+s) p_{i} + r q_{i} )$.
\end{enumerate}

Finally, we find
\eq
\label{pts}
\begin{array}{rcl}
\sigma(h(a,m),c(m)) &=& \frac{1}{4}{\displaystyle\prod_{n=1}^{\infty}}
                      ( 1 - t^n )^{-1} \\
&& ( 4 +  2 {\displaystyle\sum_{\{(p_1 , q_1 ), \ldots , (p_r , q_r )\}
   \in I_1}} (-2)^r
                   {\displaystyle\prod_{i=1}^r } \eps^+ (m, p_i , q_i )
                    t^{ p_i q_i } \\
&&      - t^{pq} {\displaystyle\sum_{\{(p_1 , q_1 ), \ldots , (p_r , q_r )\}
 \in I_2}}
       (-2)^r       {\displaystyle\prod_{i=1}^r } \eps^+ (m, p_i , q_i )
                    t^{ p_i q_i } \\
&&    + 2 {\displaystyle\sum_{ \{ (p_1 , q_1 ), \ldots ,
      (p_r , q_r )\} \in I_3 }} (-2)^r
                   {\displaystyle\prod_{i=1}^r } \eps^- (m, p_i , q_i )
                    t^{p_i q_i } \\
&&    + t ^{pq} {\displaystyle\sum_{\{ (p_1 , q_1 ), \ldots , (p_r , q_r )\}
    \in I_4 }}
    (-2)^r    {\displaystyle\prod_{i=1}^r} \eps^- (m, p_i , q_i )
                    t^{ p_i q_i } ) \, .
\end{array}
\en
Here $p$ and $q$ are positive integers with $(r+s)p - r q = -a$, and the
product $pq$ is minimized subject to these conditions.
The sums are over finite sequences of pairs of positive integers, subject
to the following conditions.
The set $I_1$ consists of sequences such that
\begin{enumerate}
\item $| (r+s) p_1 - r q_1 | > a$ or \newline $(r+s) p_1 - r q_1 = -a$,
\item $| (r+s) p_{i+1} - r q_{i+1} | > (r+s) p_{i} + r q_{i}$
or \newline $(r+s) p_{i+1} - r q_{i+1} = - ( (r+s) p_{i} + r q_{i} )$;
\end{enumerate}
$I_2$ consists of sequences such that
\begin{enumerate}
\item $| (r+s) p_1 - r q_1 | > (r+s)p + rq $ or \newline
$(r+s) p_1 - r q_1 = -((r+s)p + rq )$,
\item $| (r+s) p_{i+1} - r q_{i+1} | > (r+s) p_{i} + r q_{i}$
or \newline $(r+s) p_{i+1} - r q_{i+1} = - ( (r+s) p_{i} + r q_{i} )$;
\end{enumerate}
$I_3$ consists of sequences such that
\begin{enumerate}
\item $| (r+s) p_1 - r q_1 | > a$ or \newline $(r+s) p_1 - r q_1 = a$,
\item $| (r+s) p_{i+1} - r q_{i+1} | > (r+s) p_{i} + r q_{i}$
or \newline $(r+s) p_{i+1} - r q_{i+1} =  ( (r+s) p_{i} + r q_{i} )$;
\end{enumerate}
$I_4$ consists of sequences such that
\begin{enumerate}
\item $| (r+s) p_1 - r q_1 | > (r+s)p + rq$ or \newline
   $(r+s) p_1 - r q_1 = (r+s)p + rq$,
\item $| (r+s) p_{i+1} - r q_{i+1} | > (r+s) p_{i} + r q_{i}$
or \newline $(r+s) p_{i+1} - r q_{i+1} =  ( (r+s) p_{i} + r q_{i} )$.
\end{enumerate}

Note that, when $W(h,c)$ is unitary,
\eq
\label{chars}
\sigma ( h,c) ~=~ \sum_{n=0}^{\infty} {\rm rank}( M_n ) \, t^n ~=~
t^{-h} \chi(h,c) \, ,
\en
where $\chi (h,c)$ is the (standard) character of $W(h,c)$.
Thus equations (\ref{pts},\ref{chars}), applied to the discrete series
of unitary
representations, imply a set of rather complicated partition identities.

In closing, we note that,
although no definite application has yet been found, it has often been
speculated that the Virasoro signature characters should play a r\^{o}le in
the physics, or the classification, of non-unitary conformal field
theories.\cite{zubetal}  It should be possible to test this hypothesis, once
the modular properties of expressions (\ref{gaps}) and (\ref{pts})
are understood.

\acknowledgement

I am very grateful to D. Friedan and S. Shenker for many discussions of
the FQS theorem and the signature character problem, to
G. Watts for a related collaboration\cite{akgw} and other help,
to P. Goddard for helpful conversations, and to
J.-B. Zuber for pointing out errors in a draft.
This work was supported in part by funds from the US National Science
Foundation under grant number PHY89-04035, by an SERC Advanced
Fellowship and by the Knox-Shaw Research Fellowship at Sidney Sussex
College, Cambridge. The hospitality of the
Santa Barbara Institute for Theoretical Physics is acknowledged with thanks.

\end{document}